\documentclass[12pt]{article}
\usepackage{amsmath,amsfonts}
\usepackage[titletoc,title]{appendix}
\usepackage{pdfsync}
\usepackage{verbatim}
\usepackage{amsmath,amsfonts,amssymb}
\usepackage{amsthm}
\theoremstyle{definition}
\newtheorem{example}{Example}[]
\usepackage{stmaryrd}
\usepackage{natbib}
\usepackage{graphicx}
\usepackage{rotating}
\usepackage[T1]{fontenc}
\def \grad{\mbox{grad\hskip 1pt}}
\def \Grad{\mbox{Grad\hskip 1pt}}

\begin{document}
\title{\bf On deformation-gradient tensors as two-point tensors in curvilinear coordinates}
\author{Andrey Melnikov%
\footnote{Department of Earth Sciences, Memorial University of Newfoundland,  Canada;
{\tt amelnikov@mun.ca}}\,,
Michael A. Slawinski%
\footnote{Department of Earth Sciences,
Memorial University of Newfoundland, Canada;
{\tt mslawins@mac.com}}}
\maketitle
\begin{abstract}
We derive a general expression for the deformation-gradient tensor by invoking the standard definition of a gradient of a vector field in curvilinear coordinates.
This expression shows the connection between the standard definition of a gradient of a vector field and the deformation gradient tensor in continuum mechanics.
We illustrate its application in the context of problems discussed by \citet{Ogden}.
\end{abstract}
%%%%%%%%%%%%%%%%%%%%%%%%%%%%%%%%%%%
\section{Gradient of vector field}
\label{sec:GradVecField}
%%%%%%%%%%%%%%%%%%%%%%%%%%%%%%%%%%%
Let us consider a gradient of a vector field in the curvilinear coordinates in~${\mathbb R}^3$\,,
\begin{equation}\label{eq:start-grad}
  \grad {\bf v} :=\sum_{j=1}^3\frac{\partial{\bf v}}{\partial x^j}\otimes{\bf g}^j(\bf x)\,,
\end{equation}
where $\bf x$ is a point in space~${\mathbb R}^3$.
If we write ${\bf v}=\sum_{k=1}^3v^k\,{\bf g}_k(\bf x)$\,, expression~\eqref{eq:start-grad} becomes
\begin{equation}\label{eq:prod-rule}
\grad {\bf v}
=
\sum_{k=1}^3\sum_{j=1}^3\frac{\partial v^k}{\partial x^j}{\bf g}_k{\bf (x)}\otimes{\bf g}^j{\bf (x)} +
\sum_{k=1}^3\sum_{j=1}^3v^k\frac{\partial {\bf g}_k{\bf (x)}}{\partial x^j}\otimes{\bf g}^j{\bf (x)}
\,.
\end{equation}
Since
\begin{equation}\label{eq:Christoffel}
\frac{\partial{\bf g}_k}{\partial x^j}=\sum_{m=1}^{3}\Gamma^m_{kj}\,{\bf g}_m(\bf x)\,,
\end{equation}
where $\Gamma^m_{kj}$ is the Christoffel symbol, expression~\eqref{eq:prod-rule} can be rewritten as
\begin{equation*}\label{withChristoffel}
\grad {\bf v}=\sum_{k=1}^3\sum_{j=1}^3\frac{\partial v^k}{\partial x^j}{\bf g}_k{\bf (x)}\otimes{\bf g}^j{\bf (x)}+\sum_{m=1}^{3}\sum_{k=1}^{3}\sum_{j=1}^{3}\Gamma^m_{kj}v^k{\bf g}_m({\bf x})\otimes{\bf g}^j{\bf (x)}\,.
\end{equation*}
Exchanging in the second term the summation indices, $m$ and $k$\,, we obtain
\begin{equation}\label{eq:final}
\grad {\bf v}=\sum_{m=1}^{3}\sum_{k=1}^{3}\sum_{j=1}^{3}\left(\frac{\partial v^k}{\partial x^j}+\Gamma^k_{mj}v^m\right){\bf g}_k({\bf x})\otimes{\bf g}^j{\bf (x)}\,.
\end{equation}
Expression~\eqref{eq:final} is known and appears, for instance, in \citet[p.~60]{Ogden}.
%%%%%%%%%%%%%%%%%%%%%%%%%%%%%%%%%%%
\section{Deformation-gradient tensor}
\label{sec:FinCurviLin}%TS
%%%%%%%%%%%%%%%%%%%%%%%%%%%%%%%%%%%
In the absence of mechanical loads, a body occupies a configuration,~$\mathcal{B}_r$\,, whose boundaries are $\partial\mathcal{B}_r$\,.
Herein, $r$ denotes the reference configuration.
Upon application of a load, the configuration of the body changes; the new configuration, which is referred to as the current one, is denoted by $\mathcal{B}$\,, and its boundary by~$\partial\mathcal{B}$\,.
The material points---in the reference and the current configurations---are denoted by the corresponding position vectors, $\bf X$ and $\bf v$\,, respectively.
Deformation is described by the vector function $\boldsymbol{\chi}$ such that $\bf v=\boldsymbol{\chi}(\bf X)$.

Deformation gradient tensor, as a two-point tensor, is defined by
\begin{equation}\label{eq:startGrad}
\Grad {\bf v}:=\sum_{j=1}^{3}\frac{\partial {\bf v}}{\partial X^j}\otimes{\bf G}^j({\bf X})\,.
\end{equation}
Examining expressions~\eqref{eq:start-grad} and \eqref{eq:startGrad}, we see that the former can be viewed as the gradient of a vector field,~$\bf v$\,, in~$\mathcal{B}$\,.

In a manner analogous to the one presented in Section~\ref{sec:GradVecField}, we write
\begin{equation*}\label{eq:2pointprod-rule}
\Grad {\bf v}=\sum_{k=1}^{3}\sum_{j=1}^{3}\frac{\partial v^k}{\partial X^j}{\bf g}_k({\bf x})\otimes{\bf G}^j({\bf X})+\sum_{k=1}^{3}\sum_{j=1}^{3}v^k\frac{\partial {\bf g}_k(\bf x)}{\partial X^j}\otimes{\bf G}^j({\bf X})\,,
\end{equation*}
which, following the chain rule, becomes
\begin{equation*}\label{eq:2point-chain-rule}
\Grad {\bf v}=\sum_{k=1}^{3}\sum_{j=1}^{3}\frac{\partial v^k}{\partial X^j}{\bf g}_k({\bf x})\otimes{\bf G}^j({\bf X})+\sum_{k=1}^{3}\sum_{m=1}^{3}\sum_{j=1}^{3} v^k\frac{\partial {\bf g}_k(\bf x)}{\partial x^m}\frac{\partial x^m}{\partial X^j}\otimes{\bf G}^j({\bf X})\,.
\end{equation*}
Using expression~\eqref{eq:Christoffel}, we write
\begin{equation*}\label{eq:withChristoffel}
\Grad {\bf v}=\sum_{k=1}^{3}\sum_{j=1}^{3}\frac{\partial v^k}{\partial X^j}{\bf g}_k({\bf x})\otimes{\bf G}^j({\bf X})+\sum_{k=1}^{3}\sum_{d=1}^{3}\sum_{m=1}^{3}\sum_{j=1}^{3}v^k\Gamma^d_{km}{\bf g}_d({\bf x})\frac{\partial x^m}{\partial X^j}\otimes{\bf G}^j({\bf X})\,.
\end{equation*}
Exchanging in the second term the summation indices, $d$ and $k$\,, we write
\begin{equation}\label{eq:2point-final}
\Grad {\bf v}=\sum_{k=1}^{3}\sum_{d=1}^{3}\sum_{m=1}^{3}\sum_{j=1}^{3}  \left(\frac{\partial v^k}{\partial X^j}+v^d\Gamma^k_{dm}\frac{\partial x^m}{\partial X^j}\right){\bf g}_k({\bf x})\otimes{\bf G}^j(\bf X)\,.
\end{equation}
This is the sought expression of the deformation gradient, which is a two-point tensor.
It is a two-point counterpart of expression~\eqref{eq:final}.

In expressions~\eqref{eq:startGrad} and \eqref{eq:2point-final}, a position vector,~$\bf v$\,, and a point,~$\bf x$\,, constitute the same entity, namely, a material point in the current configuration.
Nonetheless, it is preferable to make a distinction in notation.
Otherwise, a contravariant component,~$v^k$\,, may be confused with a curvilinear coordinate $x^k$\,.
For instance, in spherical coordinates, $\bf x$\,, which is expressed in terms of $x^1=r$\,, $x^2=\theta$ and $x^3=\phi$\,, can be identified in terms of a single component, ${\bf v}=r\,{\bf e}_r$\,, with respect to the basis vector, ${\bf g}_1={\bf e}_r$\,, since ${\bf e}_r$ depends on both $\theta$ and $\phi$\,.
Thus, for a given point in space, there is no direct correspondence between the curvilinear coordinates and the components of the position vector.
In contrast, in expression~\eqref{eq:Grad-vector-field}, $\bf v$ and $\bf x$ refer to distinct entities.

In expression~\eqref{eq:2point-final}, $\partial v^k/\partial X^j$ is a partial derivative of a contravariant component,~$v^k$\,, with respect to $X^j$\, in the reference configuration; $\partial x^m/\partial X^j$ is partial derivative of a coordinate,~$x^m$\,, of a material point in the current configuration, with respect to its curvilinear coordinate $X^j$ in the reference configuration.
Application of expression~\eqref{eq:2point-final} to specific problems is illustrated in Section~\ref{sec:SpecCases}.

In contrast to expressions~\eqref{eq:startGrad} and~\eqref{eq:2point-final}, where $\bf v$ represents the position of a material point in the current configuration, let us consider the vector field,~$\bf v$\,, in the current configuration, with the positions of material points depending on positions in the reference configuration.
In such a case, expression~\eqref{eq:2point-final} becomes
\begin{equation}\label{eq:Grad-vector-field}
\Grad {\bf v}=\sum_{p=1}^{3}\sum_{k=1}^{3}\sum_{d=1}^{3}\sum_{m=1}^{3}\sum_{j=1}^{3}  \left(\frac{\partial v^k}{\partial x^p}\frac{\partial x^p}{\partial X^j}+v^d\Gamma^k_{dm}\frac{\partial x^m}{\partial X^j}\right){\bf g}_k({\bf x})\otimes{\bf G}^j(\bf X)\,.
\end{equation}
%%%%%%%%%%%%%%%%%%%%%%%%%%%%%%%%%%%
\section{Specific cases}
\label{sec:SpecCases}
%%%%%%%%%%%%%%%%%%%%%%%%%%%%%%%%%%%
Let us apply expression~\eqref{eq:2point-final} to various deformations, and obtain the corresponding deformation gradient tensors.
%%%%%%%%%%%%%%%%%%%%%%%%%%%%%%%%%%%
\subsection{Spherical symmetry}
%%%%%%%%%%%%%%%%%%%%%%%%%%%%%%%%%%%
Let us consider the problem of inflation of a thick-walled spherical shell, described in
 \citet[p.~116]{Ogden}.
Spherically symmetric deformation is
\begin{equation}\label{eq:spher-deform}
{\bf v}=f(R){\bf X}\,,
\end{equation}
where $R=\|{\bf X}\|$\,.
Since in the reference configuration
\begin{equation*}
{\bf X}=R\,{\bf E}_R\,,
\end{equation*}
relation~\eqref{eq:spher-deform} can be written as
\begin{equation}\label{eq:connection-x-X}
{\bf v}=f(R)\,R\,{\bf E}_R=r\,{\bf e}_r\,,
\end{equation}
where, due to the nature of deformation under consideration, ${\bf E}_R\equiv{\bf e}_r$\,.
Therefore, such a deformation can be described in terms of spherical coordinates---in the reference and current configurations---by using
\begin{equation}\label{eq:polar-coord}
r=f(R)R\,, \quad \theta=\Theta\,, \quad \phi=\Phi\,.
\end{equation}
For incompressible materials
$f(R)$ has a specific form \citet[p.~371]{Slawinski}.

In expression~\eqref{eq:connection-x-X} the components of $\bf v$  are stated in terms of spherical coordinates; in expression~\eqref{eq:2point-final} the components of $\bf v$ are stated with respect to the curvilinear basis,  ${\bf g}_k(\bf x)$\,, $k\in\{1,2,3\}$\,.
Therefore, we use connections
\begin{align}\label{eq:physical-conrovar-comp}
  v^1 & =v_r\,, \\
  v^2 & =\frac{v_{\theta}}{r}\,, \\
  v^3 & =\frac{v_{\phi}}{r \sin\theta}\,.
\end{align}
Herein, due to expression~\eqref{eq:connection-x-X}, we have $v^1 = v_r=r=f(R)\,R$\,; the other components are zero.
For spherical coordinates the only nonzero Christoffel symbols are
\begin{align}\label{eq:Christoffel values}
  \Gamma^2_{12} & =\Gamma^3_{13}=\frac{1}{r}\,, \quad \Gamma^1_{22}=-r\,, \quad \Gamma^3_{23}=\cot \theta\,,\notag \\[1ex]
  \Gamma^1_{33} & =-r\sin^2\theta\,, \quad \Gamma^2_{33}=-\sin\theta\cos\theta\,.
\end{align}
It follows that
\begin{align}\label{eq:gsmall}
  {\bf g}_1(\bf x) & = {\bf e}_r\,, \\
  {\bf g}_2(\bf x) & = r{\bf e}_{\theta}\,,\\
  {\bf g}_3(\bf x) & = r\sin\theta\,{\bf e}_\phi\,.
\end{align}
Also,
\begin{align}
  {\bf G}^1(\bf X) & = {\bf E}_R\label{eq:spher-Gbig}\,, \\
  {\bf G}^2(\bf X) & = \frac{{\bf E}_{\Theta}}{R}\,,\\[1ex]
  {\bf G}^3(\bf X) & = \frac{{\bf E}_\Phi}{R\sin\Theta}\,. \label{eq:Gbig}
\end{align}
Using expressions~\eqref{eq:polar-coord}--\eqref{eq:Gbig} in expression~\eqref{eq:2point-final}, we obtain
\begin{equation*}
\Grad{\bf v}=\left[f'(R)R+f(R)\right]{\bf e}_r\otimes{\bf E}_R+f(R){\bf e}_{\theta}\otimes{\bf E}_{\Theta}+f(R){\bf e}_{\phi}\otimes{\bf E}_{\Phi}\,.
\end{equation*}
%%%%%%%%%%%%%%%%%%%%%%%%%%%%%%%%%%%
\subsection{Cylindrical symmetry}
%%%%%%%%%%%%%%%%%%%%%%%%%%%%%%%%%%%
Let us consider three examples that involve a cylindrical symmetry.
To use expression~\eqref{eq:2point-final}, we invoke
\begin{align}\label{eq:cyl-physical-controvar-comp}
  v^1 & =v_r\,, \\
  v^2 & =\frac{v_{\theta}}{r}\,, \\
  v^3 & =v_z\,.
\end{align}
The only nonzero Christoffel symbols are
\begin{equation}\label{eq:cyl-Christoffel values}
\Gamma^2_{21} =\Gamma^2_{12}=\frac{1}{r}\,, \quad \Gamma^1_{22}=-r\,.
\end{equation}
Also,
\begin{align}\label{eq:cyl-gsmall}
  {\bf g}_1(\bf x) & = {\bf e}_r\,, \\
  {\bf g}_2(\bf x) & = r\,{\bf e}_{\theta}\,,\\
  {\bf g}_3(\bf x) & = {\bf e}_z\,.
\end{align}
Furthermore,
\begin{align}
  {\bf G}^1(\bf X) & = {\bf E}_R\,, \label{eq:cyl-Gbigf}\\
  {\bf G}^2(\bf X) & = \frac{{\bf E}_{\Theta}}{R}\,,\\[1ex]
  {\bf G}^3(\bf X) & = {\bf E}_Z\,. \label{eq:cyl-Gbig}
\end{align}
\begin{example}
Let us consider the problem of combined extension and torsion of a cylinder, described by \citet[p.~112]{Ogden}.
In the reference configuration, a cylindrical body can be described in terms of cylindrical coordinates $(R, \theta, Z)$ by inequalities
\begin{equation*}
0\leqslant R\leqslant A\,, \quad  0\leqslant \Theta\leqslant 2\pi\,, \quad  0\leqslant Z\leqslant L\,,
\end{equation*}
where $A$ and $L$ are the radius and the length of the cylinder.
The cylindrical body is extended uniformly so that in terms of cylindrical coordinates it can be described as
\begin{equation*}
0\leqslant r\leqslant a\,, \quad  0\leqslant \theta\leqslant 2\pi\,, \quad  0\leqslant z\leqslant\ell\,,
\end{equation*}
where $\ell$ is the length in the current configuration, given by $\ell=\lambda L$\,, and $a$ is the radius in the current configuration, given by $a=\lambda^{-1/2}A$\,.

The deformation is described by
\begin{equation*}
r=\lambda^{-1/2}R\,, \quad \theta=\Theta\,, \quad  z=\lambda Z\,.
\end{equation*}

Then the plane face at $z=\ell$ is rotated, while the face at $z=0$ remains fixed.
The rotating planes remain normal to the axis of the cylinder.
Radius $r$ is rotated by the angle $\tau z$, where $\tau$ represents the twist per unit length of $z$ in the deformed configuration.

The total deformation is given by
\begin{equation*}\label{eq:cyl-coord}
r=\lambda^{-1/2}R\,, \quad \theta=\Theta+\tau\lambda Z\,, \quad z=\lambda Z\,.
\end{equation*}
The position vector in the reference and current configurations can be written as
\begin{equation}
\label{eq:ext-torsion}
{\bf X}=R{\bf E}_{R}+Z{\bf E}_Z\,, \quad {\bf v}=\lambda^{-1/2}R\,{\bf e}_r+\lambda Z\,{\bf e}_z\,.
\end{equation}
Substitution of expressions~\eqref{eq:cyl-physical-controvar-comp}--\eqref{eq:ext-torsion} into expression~\eqref{eq:2point-final} results in
\begin{equation}
\label{eq:defor-extension-torsion}
\Grad{\bf v}=\lambda^{-1/2}\,{\bf e}_r\otimes{\bf E}_R+\lambda^{-1/2}\,{\bf e}_{\theta}\otimes{\bf E}_{\Theta}+\lambda\,{\bf e}_{z}\otimes{\bf E}_{Z}+\tau r \lambda \,{\bf e}_{\theta}\otimes{\bf E}_Z\,.
\end{equation}
This expression can be decomposed into a local rotation, simple extension and a simple shear \citep[p.~113]{Ogden}.

Let us comment on components of tensor~\eqref{eq:defor-extension-torsion}.
Entry~$11$ of its matrix expression is
\begin{equation*}
\Big(\underbrace{\frac{\partial v^1}{\partial X^1}}_{\text{$\lambda^{-1/2}$}} +\underbrace{\sum_{d=1}^{3}\sum_{m=1}^{3}v^d \Gamma^1_{dm}\frac{\partial x^m}{\partial X^1}}_{\text{$0$}}\Big)\,;
\end{equation*}
entry~$23$ is
\begin{equation*}
\Big(\underbrace{\frac{\partial v^2}{\partial X^3}}_{\text{$0$}} +\underbrace{v^1}_{r} \underbrace{\Gamma^2_{12}}_{1/r}\underbrace{\frac{\partial x^2}{\partial X^3}}_{\partial \theta/\partial Z}\Big) r\,.
\end{equation*}
\end{example}
\begin{example}
Let us consider the combination of axial shear and torsion of a cylindrical tube defined by expressions
\begin{equation*}
r=R\,, \quad \theta=\Theta+\tau\left[Z+\omega(R)\right]\,, \quad z=Z+\omega(R)\,,
\end{equation*}
given in \citet[p.~117]{Ogden}.
Thus, the position vector in the current configuration is
\begin{equation*}
{\bf v}=r{\bf e}_r+\left[Z+\omega(R)\right]{\bf e}_z\,.
\end{equation*}
According to expression~\eqref{eq:2point-final},
\begin{align}\label{eq:axial-torsion-Grad}
\Grad{\bf v}=\, & {\bf e}_r\otimes {\bf E}_R +  {\bf e}_{\theta}\otimes {\bf E}_{\Theta} +  {\bf e}_z\otimes {\bf E}_Z \nonumber\\
&+ \omega'(R)\,{\bf e}_z\otimes{\bf E}_R+\tau r \omega'(R)\,{\bf e}_{\theta}\otimes{\bf E}_R+\tau r\,{\bf e}_{\theta}\otimes {\bf E}_Z\,.
\end{align}
Since the position vector, ${\bf v}$\,, is expressed with respect to ${\bf e}_r$ and ${\bf e}_z$\,, we can write expression~\eqref{eq:2point-final} in terms of cylindrical coordinates and proceed as follows.
Inserting expressions~\eqref{eq:cyl-Gbigf}--\eqref{eq:cyl-Gbig} into expression~\eqref{eq:startGrad}, we obtain
\begin{equation}\label{eq:Grad-cyl}
  \Grad {\bf v}=\frac{\partial {\bf v}}{\partial R}\otimes {\bf E}_R+\frac{1}{R}\frac{\partial \bf v}{\partial \Theta}\otimes{\bf E}_{\Theta}+\frac{\partial \bf v}{\partial Z}\otimes {\bf E}_Z\,,
\end{equation}
where
\begin{eqnarray}\label{eq:cyl-dv/dR}
\frac{\partial \bf v}{\partial R} &=& {\bf e}_r+r \frac{\partial {\bf e}_r}{\partial \theta}\frac{\partial \theta}{\partial R}+\omega'(R){\bf e}_z\nonumber \\[1ex]
&=&{\bf e}_r+r \tau \omega'(R) {\bf e}_{\theta}+\omega'(R){\bf e}_z\,,
\end{eqnarray}
\begin{equation}
\frac{\partial \bf v}{\partial \Theta} = R \frac{\partial {\bf e}_r}{\partial \theta}\frac{\partial \theta}{\partial \Theta}=R{\bf e}_{\theta}\,,
\end{equation}
\begin{equation}\label{eq:cyl-dv/dZ}
\frac{\partial \bf v}{\partial Z} = R \frac{\partial {\bf e}_r}{\partial \theta}\frac{\partial \theta}{\partial Z}+{\bf e}_z=R\tau{\bf e}_{\theta}+{\bf e}_z\,.
\end{equation}
Hence, inserting expressions~\eqref{eq:cyl-dv/dR}--\eqref{eq:cyl-dv/dZ} into expression~\eqref{eq:Grad-cyl} results in expression~\eqref{eq:axial-torsion-Grad}.

If we insert expressions~\eqref{eq:spher-Gbig}--\eqref{eq:Gbig} into expression~\eqref{eq:startGrad}, we obtain an expression, which may be used for problems with a spherical symmetry
\begin{equation*}
\Grad{\bf v}=\frac{\partial{\bf v}}{\partial R}\otimes{\bf E}_R+\frac{1}{R}\frac{\partial{\bf v}}{\partial \Theta}\otimes{\bf E}_\Theta+\frac{1}{R \sin \Theta}\frac{\partial {\bf v}}{\partial \Phi}\otimes{\bf E}_{\Phi}\,.
\end{equation*}
\end{example}
\begin{example}
Let us consider a combined axial and azimuthal shear of a circular tube, discussed by \citet[p.~115]{Ogden}.
This type of deformation is defined by
\begin{equation*}
r=R\,, \quad \theta=\Theta+\phi(R)\,, \quad  z=Z+\omega(R)\,.
\end{equation*}
The position vector in the current configuration is
\begin{equation*}
{\bf v}= r\,{\bf e}_r + [Z+\omega(R)]\,{\bf e}_z\,.
\end{equation*}
Thus, application of expression~\eqref{eq:2point-final} results in
\begin{align}\label{eq:de-grad-axial-azimuthal}
  \Grad {\bf v} =  {\bf e}_r\otimes {\bf E}_R + {\bf e}_{\theta}\otimes {\bf E}_{\Theta} +  &{\bf e}_z\otimes {\bf E}_Z\nonumber\\
  &+ r \phi'(R) {\bf e}_{\theta}\otimes {\bf E}_R + \omega'(R) {\bf e}_z\otimes {\bf E}_R\,.
\end{align}
This expression can be decomposed into local simple shears and a rotation \citet[p.~116]{Ogden}).

In cylindrical coordinates only two components are required to state a position vector.
They are expressed with respect to ${\bf e}_r$ and ${\bf e}_z$\, in the current configuration and with respect to ${\bf E}_r$ and ${\bf E}_z$\, in the reference configuration.

In spherical coordinates only one component is required to state a position vector.
It is expressed with respect to ${\bf e}_r$\, in the current configuration  and with respect to ${\bf E}_r$\, in the reference configuration.

For a general vector field, however, the three components are required with respect to corresponding basis vectors in cylindrical and spherical coordinates.
\end{example}
%%%%%%%%%%%%%%%%%%%%%%%%%%%%
\section*{Acknowledgments}
%%%%%%%%%%%%%%%%%%%%%%%%%%%%
We wish to acknowledge insightful comments of Raymond Ogden and proofreading by David Dalton and Theodore Stanoev.
This research was performed in the context of The Geomechanics Project supported by Husky Energy.
Also, this research was partially supported by the Natural Sciences and Engineering Research Council of Canada, grant 238416-2013.
%%%%%%%%%%%%%%%%%%%%%%%%%%%%

\bibliographystyle{unsrtnat}
\bibliography{MS}
%%%%%%%%%%%%%%%%%%%%%%%%%%%%
\end{document}